\documentclass[11pt]{article}
\usepackage{color}

\usepackage{latexsym}  
\usepackage{amssymb}
\usepackage{graphicx}
\usepackage{amsmath}
\usepackage{mathrsfs}

\newcommand{\be}{\begin{equation}}
\newcommand{\ee}{\end{equation}}
\newcommand{\bea}{\begin{eqnarray}}
\newcommand{\eea}{\end{eqnarray}}
\newcommand{\bref}[1]{(\ref{#1})}

\newcommand{\pa}{\partial}

\begin{document}
\begin{titlepage}
\begin{flushright}
\today
\end{flushright}

\begin{center}
{\Large\bf  
The self-attractive ultralight axion and its time scale of collapsing in general relativistic framework }
\end{center}

\begin{center}

\vspace{0.1cm}

{\large Takeshi Fukuyama$^{a,}$%
\footnote{E-mail: fukuyama@rcnp.osaka-u.ac.jp}}

\vspace{0.2cm}

{\small \it ${}^a$Research Center for Nuclear Physics (RCNP),
Osaka University, \\Ibaraki, Osaka, 567-0047, Japan}



\end{center}

\begin{abstract}
The formation of supermassive black holes (SMBHs) at high redshift $z$ by ultralight axion dark matter (DM) is discussed in the general relativistic framework.
The critical condition of collapsing of self-attracting DM in non-relativistic treatment corresponds to dust particle,  which allows to estimate the time scale of the SMBH formation and the its length scale in the general relativistic framework. It gives the support of ultralight axion model based on the string model as the origins of the early universe SMBH and  the Little Red Dots (LRDs) etc. observed by James Webb Space Telescope (JWST).
\end{abstract}
\end{titlepage}

\section{Introduction}
The recent observation shows that the super-massive black holes (SMBHs)  are formed as early as several hundreds million years after the big  bang \cite{Richstone1998, Kormendy, Banados, JWST}, which needs the explanation since the accretion processes of both baryonic and collisionless DM are rather slow.
In the recent paper \cite{Fuku1}, we studied the reason of too-early formation of SMBHs based on the string-inspired axion model.  There, we left the important point on the time-scale of the SMBH formation open. That is, we explained the self-attraction of axion dark matter (DM) causes the early formation of SMBH. However, we have not discussed its time scale concretely.  This is due to the fact that our formulation in \cite{Fuku1} was essentially non relativistic and needs general relativistic treatment for the estimation of the time scale of the formation of SMBH, which is the purpose of the present letter.
This makes possible to relate our model with the recent observations of the Little Red Dots (LRDs) \cite{Harikane, Finkelstein, Matthee, Maiolino} from the following reasons \cite{Li2024, Labbe2023}:

1. More than 70\% of LRDs exhibit broad hydrogen Balmer lines, indicating the presence of massive BHs.

2. The abundance in cosmic volume is significantly higher than what is expected so far.

3. BH to galaxy mass ratio is much larger than that of neaby galaxies.

Though we consider explicitly DM only since it dominates the dynamics, it is consistent with the above mentioned properties.

This paper is organized as follows.
In Sec.2 we will review on the non-relativistic formation of SMBH \cite{Fuku1}. It is emphasized that the critical point of collapsing is noting but the total pressure=0 by virtue of the self-attraction of axion DM.  In Sec.3 the general relativistic estimation of the time scale of SMBH formation and their size will be given. Section 4 is devoted to discussions.  In Appendix, we resume the equation of state of the usual matters (Bosons and Fermions) to illustrate the peculiarity of axion DM.

In this letter, we use $\hbar=c=1$ units.
\section{Review of SMBH formation due to ultralight axion DM}
We consider the axion DM \cite{PQ, Marsh2016, Soda2020} in the big bang universe, where the metric $g_{\mu\nu}$ is
\be
ds^2=(1+2\Phi)dt^2-a^2(1-2\Phi)(dx^2+dy^2+dz^2).
\label{metric}
\ee
The effective action of axion field $\phi$ is given by
\bea
\label{action}
S&=&\int d^4x \sqrt{-g}[\frac{1}{2}R-\frac{1}{2}g^{\mu\nu}\partial_\mu\phi\partial_\nu\phi-\Lambda^4\left(1-\cos\frac{\phi}{f_a}\right)\nonumber\\
&+&\sum_iy_i\frac{\partial_\mu\phi}{f_a}\overline{\psi}_i\gamma^\mu\gamma_5\psi_i 
\eea
Substituting \bref{metric} into \bref{action}, we obtain
\be
\ddot{\phi}+3H\dot{\phi}-4\dot{\Phi}\dot{\phi}-\frac{1+4\Phi}{a^2}\Delta\phi+m_a^2(1+2\Phi)\phi+\frac{\lambda}{3!}\phi^3=0.
\label{phieq}
\ee
Here $\dot{\phi}\equiv \frac{\pa\phi}{\pa t}$ and $H=\frac{\dot{a}}{a}$, and
\be
m_a^2=\frac{\Lambda^4}{f_a^2},~~\lambda=-\frac{\Lambda^4}{f_a^4}
\label{ma}
\ee
and we have neglected the coupling with the ordinary matters.
We are concerned with the string-inspired axion, where $\Lambda^4$ is given by
\be
\Lambda^4=\Lambda_{string}^4=M_{SUSY}^2M_{Pl}^{*2}e^{-S_{instanton}}.
\label{SUSY}
\ee
Here 
\be
S_{instanton}=\frac{1}{4g^2}\int d^4xTr{F_{\mu\nu}\tilde{F}^{\mu\nu}}
\label{instanton2}
\ee
with $F_{\mu\nu}\equiv F_{\mu\nu}^a\lambda^a$, and
$f_a$ and $m_a$ become independent parameters unlike the QCD case and they are given as follow:  Firstly, $f_a$ is \cite{Witten}
\be
f_a=\frac{\alpha_{GUT}M_{Pl}^*}{\sqrt{2}2\pi}\approx 1.1\times 10^{16}~\mbox{GeV}.
\label{GUT}
\ee
Here we have set the strong coupling constant $\alpha_{GUT}\equiv \frac{g^2}{4\pi}$ as the unified coupling constant at GUT scale,
\be
\alpha_{GUT}=\frac{1}{25}.
\label{aGUT}
\ee
The string-inspired axion mass is
\be
m_a=\frac{M_{SUSY}M_{Pl}^*}{f_a}e^{-\frac{S_{instanton}}{2}}
\label{SUSYmass}
\ee
with supersymmetry breaking scale $M_{SUSY}$. 
However, there are many parameters in axiverse \cite{Arvanitaki}, $M_{SUSY},~S_{instanton},~\mbox{misalignment parameter}~\theta_i,$
primordial isocurvature fraction etc. Visinelli and Vagnozzi obtained 
\be
S_{instanton}=198\pm 28~~ \mbox{and} ~~M_{SUSY}=10^{11} \mbox{GeV}
\label{instanton}
\ee
by Bayesian parameter inference in light of many cosmological data \cite{Visinelli}. 
Then, if we adopt the center value $198$ for $S_{instanton}$, we obtain
\be
\mbox{log}_{10}(m_a/\mbox{eV})=-21.5^{+1.3}_{-2.3}
\label{ma2}
\ee
and 
\be
M_{BH}=2.5~\tilde{l}^2\times 10^8M_\odot
\label{SMBH}
\ee
as will be shown in \bref{KaupBH}.
This may explain the origin of SMBH \cite{Richstone1998, Kormendy, Banados}.
It is very interesting that these values lead to the observed magnitude of DM
\be
\Omega_a\equiv \frac{\rho_a}{\rho_{critical}}=\frac{1}{3H_0^2M_{Pl}^{*2}}\rho_a=\frac{1}{3H_0^2M_{Pl}^{*2}}\times a_{osc}^3\times\frac{1}{2}m_a^2f_a^2
=\sqrt{\frac{m_a}{10^{-27}\text{eV}}}\left(\frac{f_a}{M_{Pl}^*}\right)^2
\label{Omega}
\ee
with the present Hubble constant $H_0(=10^{-33}$ eV).
Here in the last equality we have used axion with $m_a=10^{-21}$ eV begins to oscillate at $z=10^7$.
Thus $\Omega_a$ is within the observed value $\Omega_a=O(1)$. 
As for the ultralight mass there are arguments constraining the lower limit \cite{Dala2022}.
They gave 
\be
m>3\times 10^{-19}~\mbox{eV}
\ee
from the velocity dispersions for stars. In our model, the original DM whose radius is larger than $\sigma_{Kaup}$ forms halo around the SMBH. Though DM dominates over ordinary matters, the details of star formation is out of scope of this paper. 

If we consider further the collapsing of the DM, we may consider firstly the behavior of DM decoupled with the cosmological expansion. In this case, axion field, $\phi$, is divided into fast oscillation ($e^{im_at}$) part and slow one ($\psi$) (Please do not confuse with ordinary matters defined in \bref{action}.), as
\be
\phi=\frac{1}{\sqrt{2m_a}}\left(\psi e^{-im_at}+\psi^*e^{im_at}\right).
\label{Eq5}
\ee
Substituting \bref{Eq5} into \bref{phieq}, we obtain
\be
i\dot{\psi}+i\frac{3}{2}H\psi+\frac{1}{2m_a}\Delta \psi-m_a\Phi\psi=0.
\label{Eqpsi}
\ee
Here we consider the fluctuation (the overdense region) of the Peccei-Quinn field decoupled from the cosmological expansion and use Gaussian approximation \cite{Gupta2017} with angular momentum, 
\begin{equation}
|\psi\left(t,x\right)|=\frac{1}{\sqrt{2\pi(l+1)!\sigma^3}}\left(\frac{r}{\sigma}\right)^le^{-r^2/(2\sigma^2)}Y_{lm}(\theta, \varphi).
\label{psi}
\end{equation}
From \bref{Eqpsi} with axion self coupling \bref{phieq}, the axion Lagrangian density becomes
 \be
\mathcal{L}= \frac{i}{2}\left(\psi^*\frac{\pa\psi}{\pa t}-\psi\frac{\pa\psi^*}{\pa t}\right)-\frac{1}{2m_a}\nabla\psi^*\cdot\nabla\psi-\frac{gN}{2}|\psi|^4+N|\psi(x)|^2\int\frac{Gm_a^2}{|{\bf x}-{\bf y}|}|\psi(y)|^2d^3y.
\label{NRLag}
\ee
Here $N$ is the total number of axion particles in the overdense region. Dimensional coupling $g$ is related with the above $\lambda$ by
\be
g=\frac{\lambda}{m_a^2}=\frac{4\pi a_s}{m_a},
\label{g}
\ee
where $|a_s|$ is the scattering length. 
Then we obtain the effective potential,
\be
V_{eff}=\frac{1}{2m_a\sigma^2}-\frac{\sqrt{2}}{3\pi}\frac{GNm_a^2}{\sigma}+\frac{l(l+1)}{2m_a\sigma^2}+\frac{gN}{6\sqrt{2}\pi^{3/2}\sigma^3}.
\label{potential2}
\ee
 The effective potential \bref{potential2} has the stable orbits at \cite{Eby, Hertzberg}
\be
\sigma_{min}=\frac{\frac{\tilde{l}^2}{2m_a}\pm \sqrt{\left(\frac{\tilde{l}^2}{2m_a}\right)^2+\frac{gGN^2m_a^2}{6\pi^{5/2}}}}
{\frac{\sqrt{2}}{3\pi}GNm_a^2} 
\label{min}
\ee
with 
\be
\tilde{l}^2\equiv l(l+1)+1.
\label{l}
\ee
Here it is very important that axion DM attracts to each others since $g<0$, which allows the two extrema in \bref{min}. The minus (plus) case corresponds to stable (unstable) orbit. These maximum and minimum points coalesce at 
\be
\left(\frac{\tilde{l}^2}{2m_a}\right)^2+\frac{gGN^2m_a^2}{6\pi^{5/2}}=0
\label{min2}
\ee
in \bref{min} and the stable orbit disappears. Eq. \bref{min2} indicates that the pressure of DM is cancelled with DM self-attraction. This is also crucial for the general relativistic treatment as will be discussed in the next section. Eq. \bref{min2} leads us to so called 
Kaup radius \cite{Kaup68},
\be
\sigma_{Kaup}=\frac{\sqrt{3}}{2\pi^{1/4}}\frac{\sqrt{-g}}{\sqrt{G}m_a}=\sqrt{-6\pi^{1/2}g}\frac{M_{Pl}^*}{m_a}
\label{Kauprad}
\ee
with the reduced Planck mass $M_{Pl}^*=\sqrt{\frac{1}{8\pi G}}=2.4\times 10^{18}$ Gev and Kaup mass,
\be
M_{Kaup}=Nm_a=\frac{\sqrt{6\pi^{5/2}}}{2}\frac{\tilde{l}^2}{\sqrt{G|\lambda |}}
=25\tilde{l}^2\frac{M_{Pl}^*}{\sqrt{|\lambda|}}.
\label{Kmass}
\ee
This mass is eventually reduced to BH mass since there is no repulsive force to prevent the collapse,
\be
M_{Kaup}=M_{BH}.
\label{KaupBH}
\ee
Then, substituting the values of \bref{instanton} into \bref{Kmass}, we obtain \bref{SMBH}.
Here we have some comments on the well established axion star.
For the typical model of axion star we may refer \cite{Helfer16}.
They also considered the self attraction term and considered the Mathew equation and its critical point.
Whereas we have used the corresponding equation (Eq. (19)) as setting the critical point of instability, and after that we will use the full Einstein equation (or Tolman-Oppenheimer equation) as will be discussed in the next section. Their value of $M_{BH}$ is larger than ours if we set $\tilde{l}^2=1$.

\section{Relativistic treatment of the gravitational collapse of axion DM}
In the previous section, we studied the formation of SMBH in the framework of non-relativistic scheme.
However, as far as we discuss SMBH, it is inevitable to treat it in the general relativistic framework. 
For that purpose, we need the equation of state of axion star in general.
Axion potential is rewritten as
\be
V(\phi)=m_a^2f_a^2\left(1-\cos\frac{\phi}{f_a}\right)-\frac{m_a^2}{2}\phi^2\approx-\frac{1}{4!}\frac{m_a^2}{f_a^2}\phi^4+\frac{1}{6!}\frac{m_a^2}{f_a^4}\phi^6+ ......
\ee
Expanding $\phi$ by \bref{psi}, we obtain the pressure due to the self-attraction \cite{Chavanis} up to the second order,
\be
p(\rho)=\frac{2\pi a_s}{m_a^3}\rho^2+\frac{64\pi^2a_s^2}{9m_a^6}\rho^3+ .....
\label{stateeq}
\ee
with $a_s<0$.
This was very useful for measuring the critical point \bref{min2} for causing the collapsing.
However it is not so useful for pursuing the collapsing further since $\rho$ gets larger and \bref{stateeq} is not applicable.
Instead, we should consider \bref{min2} as the cancellation of total pressure which remains valid during collapsing. There the first negative pressure term of \bref{stateeq} overwhelmingly dominates over the remaining higher order terms. Thus, we should consider axion DM as spherically symmetric pressureless dust particles without rotation (see Eq. \bref{rotation}).
That is, we can directly apply the Tolman's idea \cite{Tolman1}. 
Before discussing dust DM, we resume the equation of state $p=p(\rho)$ in usual matters in the appendix to illustrate the peculiarity of axion DM.

In axion DM case, $p$ due to self-attraction approximately given by the first term of \bref{stateeq} is negative, which is quite different from ordinary matters as we mentioned in appendix.
It causes the critical phase of collapsing peculiar to the self-attracting model. The critical point \bref{min2} appearing only in this model has been discussed by other peoples. However, no one has succeeded in describing the detail of the process of collapsing and the collapsing time scale in the presence of the self attraction \cite{Chavanis, Davoudiasl}.

Eq. \bref{min2} indicates that the collapsing phase occurs when the total pressure vanishes by virtue of the self attraction, namely $p=0$.

Substituting the central value of \bref{instanton} into \bref{Kauprad}, we obtain
\be
\sigma_{Kaup}\approx 4\times 10^{20} \mbox{cm}\approx 1\times 10^2 \mbox{pc}.
\label{Kauprad2}
\ee
\bref{Kauprad2} indicates the Gaussian maximum point and we may expect ``the radius''  may be larger by factor $10$ than this value. This may be coincident with the size of the Little Red Dots observed by JWST \cite{Harikane, Maiolino} whose median radius is $150$ pc. 

Let us proceed to discuss our dust case in general relativistic framework and find the scale of the time and length of the collapsing object as a whole.
From \bref{min2} dust gas has no effective angular momentum. This is very crucial for allowing us to use both
synchronous and comoving coordinates since in this coordinate the dust four dimensional velocity $u_\mu=\frac{dx_\mu}{ds}$ satisfies
\be
u_{\mu;\nu}-u_{\nu\mu}=\partial_\nu u_\mu-\partial_\mu u_\nu=0,
\label{rotation}
\ee
where $;$ indicates the covariant derivative. Also we must remark that the system of the outside observer
that is inertial at infinity is not sufficient for estimating the collapsing time.
Thus, we choose both synchronous and comoving coordinates \cite{LL},
\be
ds^2=d\tau^2-e^{\lambda(\tau, R)}dR^2-r^2(\tau, R)(d\theta^2+\sin^2\theta d\varphi^2).
\label{metric}
\ee
Here the coordinate system moves with matter. Then a particle characterized by the initial position $R$ 
moves with the radial velocity $\dot{r}(\tau, R)$. $r(\tau,R)$ is the radius of the particle $R$ at time $\tau$ whose circumference is $2\pi r$.
This system is decoupled with the cosmological expansion and the scale factor $a$ is not present.
$r(\tau, R)$ is determined from the Einstein'e equation,
\be
G_{\mu\nu}=8\pi GT_{\mu\nu}
\label{Einstein}
\ee.
with
\be
T^0_0=\rho, ~~T^1_1=T^2_2=T^3_3=p=0.
\label{Tmn}
\ee
Here let us compare our theory with the conventional Tolman-Oppenheimer-Volkov (TOV) equation \cite{Tolman2, Oppenheimer}.
Eq.\bref{Einstein} has three independent equations (TOV equation), whereas it includes four unknown parameters $(\lambda, \nu~(\mbox{defined by}~ e^\nu dt=d\tau), p, \mbox{and}~ \rho) $ as functions of $r$.
So we need the equation of $p=p(\rho)$. Then it is difficult to incorporate the full axion potential into this framework. However, in the collapsing phase, the equation of state correspond to \bref{min2}, $p=0$ as we mentioned, which allows the general relativistic treatment.
Though the mathematical situation is quite the same as that in Landau-Lifshitz section 103, they claimed that the dust approximation is not usually admissible in real situations.  However, by virtue of the self-attraction, we showed that the dust condition becomes real and physically admissible model. So, we obtain the time scale of collapsing as will be shown \footnote{ Though the derivations from \bref{LL1} to \bref{tau} are given in \cite{LL}, we give here since its English version has an misprint on the choice of \bref{LL7} and for self-completeness.}.

It goes from Eqs. \bref{metric} to \bref{Tmn}
\bea
\label{LL1}
 2 \dot{r}^{\prime}-\dot{\lambda} r^{\prime}&=&0 \\
 \label{LL2}
-e^{-\lambda} r^{\prime 2}+2 r \ddot{r}+\dot{r}^2+1&=&0, \\
\label{LL3}
-\frac{e^{-\lambda}}{r}\left(2 r^{\prime \prime}-r^{\prime} \lambda^{\prime}\right)+\frac{\dot{r} \dot{\lambda}}{r}+\ddot{\lambda}+\frac{\dot{\lambda}^2}{2}+\frac{2 \ddot{r}}{r}&=&0, \\
\label{LL4}
-\frac{e^{-\lambda}}{r}\left(2 r^{\prime \prime}+\frac{r^{\prime 2}}{r}-r^{\prime} \lambda^{\prime}\right)+\frac{1}{r^2}\left(r \dot{r} \dot{\lambda}+\dot{r}^2+1\right)&=&8 \pi G \rho,
\eea
\bref{LL1} is integrated as
\be
e^\lambda=\frac{r'^2}{1+f(R)},
\label{LL5}
\ee
where $f(R)$ is an arbitrary function subject to $1+f>0$. Using this \bref{LL5} and \bref{LL2}, we obtain
\be
\dot{r}^2=f(R)+\frac{F(R)}{r},
\label{LL6}
\ee
where $F(R)$ is another arbitrary function. Thus the functions $r$ and $\tau$ can be written using parameter $\eta$ and arbitrary functions $f(R)$ and $F(R)$, and an additional arbitrary function $\tau_0(R)$ as
\be
r=\frac{F}{-2f}(1-\cos\eta), ~~\tau_0(R)-\tau=\frac{F}{2(-f)^{3/2}}(\eta-\sin\eta)
\label{LL7}
\ee
for $f<0$.  It goes from \bref{LL4}, \bref{LL5} and \bref{LL6}, 
\be
8\pi G\rho=\frac{F'}{r'r^2}.
\label{LL8}
\ee
The mass within the radius $R_0$ (corresponding to $\sigma_{Kaup}$ of \bref{Kauprad}) is given by
\be
M=4\pi\int_0^{r(\tau, R_0)}\rho r^2dr=4\pi\int_0^{R_0}\rho r^2r'dR=\frac{F(R_0)}{2G}.
\label{LL9}
\ee
The last equality comes from \bref{LL7}. Thus we obtain
\be
F(R_0)=r_g.
\label{LL10}
\ee
Then we have prepared to calculate the collapsing time scale.
We need the collapsing time scale. For that purpose, we need to study the motion inside the event horizon. 
Then, as \cite{LL} indicated, we must start from a metric in vacuum that could contain both contracting and expanding space-time regions. Eq. \bref{LL7} is such a solution.
Choosing
\be
f=-\frac{1}{(R/r_g)^2+1}, ~~\tau_0=\frac{\pi}{2}r_g(-f)^{-1/2},
\ee
we get
\be
\tau=\frac{r_g}{2}\left(\frac{R^2}{r_g^2}+1\right)^{3/2}(\pi-\eta+\sin \eta),
\ee
where $0<\eta <2\pi$.
Then the world line $R=const.$ starts from $r=0$ and goes through $r=r_g$ and reaches its farthest distance  $r=r_g\left(\frac{R^2}{r_g^2}+1\right)$ at $\tau=0$ after which particle begins to fall in toward to $r_g$ and arrives once more at $r=0$ at
\be
\tau_c=\frac{\pi}{2}r_g\left(\frac{R^2}{r_g^2}+1\right)^{3/2}
\label{tau}
\ee
(See Fig.24 of Sec. 103 of \cite{LL}).  Therefore, the collapsing time scale is $\tau_c/2$.

We have insisted that the axion DM is the key ingredient of primeval SMBHs and \bref{tau} is the very important observational constraint on it. Then we impose $O(\tau_c/2=10^8)$ years as was observed by JWST and 
let us see how $\Lambda$ and $f_a$ are constrained.
\be
r_g=25\tilde{l}^2M_{Pl}^*\frac{f_a^2}{\Lambda^2}\frac{1}{M_\odot}\times 3 \mbox{Km}\approx 25\frac{M_{Pl}^*}{M_\odot}\frac{f_a^2}{\Lambda^2}\times 3 \mbox{Km}.
\label{rg2}
\ee
Here we have used the Schwarzshild radius of the solar mass $M_\odot$ is $3$ Km.
\be
R_0=\sqrt{6\pi^{1/2}|\lambda|}\frac{M_{Pl}^*}{m_a^2}=\sqrt{6\pi^{1/2}}\frac{M_{Pl}^*}{\Lambda^2}.
\label{R}
\ee
Substituting Eqs \bref{rg2} and \bref{R} into Eq. \bref{tau} with $\tau_c/2=1\times 10^8$ years, we obtain
\be
f_a\mbox{[GeV]}^4\Lambda\mbox{[GeV]}^2=1.6\times 10^{53}.
\label{tau2}
\ee
Then if we substitute the center values of \bref{instanton} into \bref{SUSY}, we obtain
\be
\Lambda=1.5 \times 10^{-7} ~\mbox{GeV}.
\label{Lambda2}
\ee
Then it goes from \bref{tau2} and \bref{Lambda2}
\be
f_a=5\times 10^{16} ~\mbox{GeV}.
\ee
This gives roughly the same value as Eq. \bref{GUT}. Also we obtain from \bref{R} and \bref{Lambda2} we obtain
\be
R_0=7.2\times 10^{18} \mbox{cm}=2.3 ~\mbox{pc}
\ee
and 
\be
M_{SMBH}\approx 5\times 10^9 M_\odot~.
\ee
Though this mass is one order of magnitude larger than the observed LRDs \cite{Harikane, Finkelstein, Matthee, Durodola} and the radius may be smaller than LRDs, they are within the error if we consider more delicate estimations.  Also it is interesting that this radius is smaller than that of \bref{Jeans} without self attraction in the nonrelativistic approach.
Thus it seems amazing that very sensitive particle data of \bref{SUSYmass} and \bref{instanton} give the roughly correct value of cosmological objects.

\section{Discussion}
We have considered ultralight string-inspired axion in the very early universe. Its self-attraction serves for axion as pressureless dust particles in the general relativistic treatment.
There, the conventional approaches use the system of the outside observer that is inertial at infinity. However it is not sufficient for studying the collapsing phases. We have rediscovered the solution which contains both contracting and expanding space-time region in the physically admissible model, and estimated the time scale of collapsing.
The difference of QCD axion and string-inspired axion in axion potential is that the former has the only one parameter $f_a$ whereas the latter has two independent parameter $f_a$ and $m_a$. In the previous calculations we fixed $m_a$ as \bref{GUT} from the axiverse \cite{Witten}. In this letter we have re-examined it from the observed time scale of the primeval SMBH as order of $10^8$ year and have obtained the same order of the value.
Also such ultralight axion may explain the nano Hz stochastic gravitational wave background \cite{JWST, Fuku1} and solve the Hubble tension \cite{Aghanim, Fuku2}.
More detailed survey of LRDs and the other very early universe objects on their total spectral distributions and AGN factors etc. will be studied in the subsequent papers.

\section*{Acknowledgments}
The author is grateful to Sai Charan Chandrasekar for useful discussions. The author also acknowledges to Professor K.S. Babu for his hospitality at the state university of Oklahoma.
This work is supported by JSPS KAKENHI Grant Numbers JP22H01237.

\appendix
\section{Equation of state in the conventional theories}
In this appendix we briefly resume the equation of state $p=p(\rho)$ in the conventional theories to illustrate the pecuriority of our theory.
We back to Eq. \bref{NRLag}, neglecting self-attraction according to the conventional approach and emphasizing this term later \cite{Soda2020}.
\be
i\dot{\psi}+i\frac{3}{2}H\psi +\frac{1}{2m_a a^2}\Delta \psi-m_a\Phi \psi=0,
\label{NR2}
\ee
where $a$ is the cosmological scale factor and $H=\frac{\dot{a}}{a}$.
Substituting $\psi=\sqrt{\frac{\rho}{m_a}}e^{i\theta}$ into \bref{NR2}, we obtain
\be
\dot{\rho}+3H\rho+\frac{1}{a}\nabla\cdot (\rho{\bf v})=0
\ee
and
\be
\frac{\partial {\bf v}}{\partial t}+H{\bf v}+\frac{1}{a}({\bf v}\cdot \nabla){\bf v}=-\frac{1}{a}\nabla P-\frac{1}{a}\nabla \Psi
\ee
from the imaginary and real parts of Eq. \bref{NR2}, respectively
with
\be
{\bf v}=\frac{1}{m_a a}\nabla\theta
\label{v}
\ee
and ``pressue'' $P$
\be
P\equiv -\frac{1}{2m_a a^2\sqrt{\rho}}\Delta\sqrt{\rho}.
\ee

Dividing $\rho$ into the background $\rho_b$ plus the fluctuation $\delta \rho$ and defining $\delta\equiv \frac{\delta\rho}{\rho_b}$, we obtain
\be
P=-\frac{1}{4m_a^2 a^2}\Delta \delta\rho=\frac{k^2}{4m_a^2a^2}\delta\rho.
\ee
The sound velocity $c_s$ is given by
\be
c_s^2=\rho\frac{\delta P}{\delta \rho}=\frac{k^2}{4m_a^2a^2}.
\ee
Thus we obtain the Jeans length $1/k_J$ from
\be
\frac{a}{k_J}=c_s\frac{1}{\sqrt{4\pi G\rho}}=\frac{k_J}{2m_a a}\frac{1}{\sqrt{4\pi G\rho}},
\label{Jeans}
\ee
\be
k_J=(16\pi Ga\rho)^{1/4}\sqrt{m_a}=(10\mbox{Kpc})^{-1}\sqrt{\frac{m_a}{10^{-21}\mbox{eV}}}.
\ee
Also we refer the equation of motion for ordinary fermion degenerate state \cite{LL2},
\be 
\mu=K\left(\frac{\rho}{m'}\right)^n,~~p=\frac{n}{n+1}K\left(\frac{\rho}{m'}\right)^{n+1},
\label{LLS}
\ee
where $\mu$ and $m'$ are the chemical potential and mass per a electron, respectively.
$K$ is a positive constant and
\be
K=(3\pi^2)^n/m' 
\ee 
with $n=2/3$ for nonrelativistic case and
\be
K=(3\pi^2)^n
\ee
with $n=1/3$ for relativistic case.
Thus we know that the self-attraction of axion field makes the quite different equation of state from those due to these conventional ones.

\end{document}